\documentclass[manuscript]{aastex}
\usepackage{graphicx,natbib}
\usepackage[backref,breaklinks,colorlinks,citecolor=blue]{hyperref}
\usepackage[all]{hypcap}

\begin{document}

\title{The Gamma-ray-emitting quasar 0202$+$149: a CSS revisited}

\author{T. An\altaffilmark{1,2,3}, Y.-Z. Cui\altaffilmark{2,1}, W.A. Baan\altaffilmark{1,4}, W.-H. Wang\altaffilmark{1,3}, P. Mohan\altaffilmark{1}}

\altaffiltext{1}{Shanghai Astronomical Observatory, Chinese Academy of Sciences, 200030 Shanghai, China. Email: antao@shao.ac.cn}
\altaffiltext{2}{School of Electrical and Electronic Engineering, Shanghai Institute of Technology, 201418, Shanghai, China.}
\altaffiltext{3}{Key Laboratory of Radio Astronomy, Chinese Academy of Sciences, 210008 Nanjing, China}
\altaffiltext{4}{Netherlands Institute for Radio Astronomy ASTRON, P.O. Box 2, 7990-AA Dwingeloo, The Netherlands}

\shorttitle{0202$+$149: a CSS revisited}
\shortauthors{An et al.}

\begin{abstract}
{PKS 0202$+$149 is a low-power radio source with blazar-like $\gamma$-ray AGN characteristics. We investigate its properties and classification in relation to its $\gamma$-ray characteristics. This source shows a hint of low frequency turnover at about 200 MHz. Radio imaging data of 0202$+$149 at different frequencies show differing morphologies on both kiloparsec (kpc) and parsec (pc) scales. The overall source shows a triple structure of a core and double lobes with a total projected size of $\sim$1.3 kpc. The compact source structure of 0202+149 is reminiscent of a compact steep spectrum (CSS) source. At pc scales a core-jet structure extends $\sim$25 pc (in projection) at a position angle perpendicular to the kpc-scale structure. The curved pc-scale structure with a jet and inner lobe suggests that the CSS nuclear activity has recently re-started although its power has been decreasing, while the kpc-scale lobes are relics of earlier activity. A maximum apparent superluminal motion of $\sim16\,c$ is detected in the jet components, indicating a highly relativistic jet flow. The brightness temperature of the core is lower than the average value found for highly-beamed, $\gamma$-ray AGNs, indicating a lower radio power and a relatively lower Doppler boosting factor. The CSS radio classification indicates that blazar-like $\gamma$-ray properties can also be manifested in low-power CSS radio sources with the appropriate jet and beaming properties.
}
\end{abstract}


%

\section{Introduction}
\label{sect:intro}

The source PKS 0202$+$149
\citep[4C $+$15.05, NRAO 91 - ][]{Romney84},
is an unusual radio source \citep[][]{PS65,Gow67} that is identified as a QSO \citep{Stickel94}.
A redshift of $z = 0.833$ based upon the [\ion{O}{3}] $\lambda$3727 and [Ne {\rm I}] $\lambda$3833 lines was estimated by \citet[][]{Stickel96}, but a more recent and smaller redshift of $z = 0.405$ \citep{Perlman98,Hea08} was reported; the latter will be used in this paper.
At intermediate resolution, PKS\,0202$+$149 has a compact triple morphology extending across $\sim$200 mas (corresponding to a projected size of $\sim$1 kpc) along a northeast-southwest direction with two relatively weaker lobe components straddling a central bright core \citep{Wang03}. At higher Very Long Baseline Interferometry (VLBI) resolution the core component itself exhibits a complicated structure with a dominant core component and a sharply bent jet terminating in an inner lobe structure or hotspot \citep{Wang03}. But no link is detected between the pc-scale and kpc-scale structures.
The kpc-scale compact structure of PKS 0202+149 may meet the definition of a compact steep spectrum source \citep{Fanti90}. 

PKS 0202$+$149
was
also identified as a $\gamma$-ray active galactic nucleus (AGN) with the Energetic Gamma Ray Experiment Telescope (EGRET) onboard the Compton Observatory \citep{Thompson95,Mattox97,Hartman99}. This has recently been confirmed by the {\it Fermi} Observatory, in the First Fermi-LAT catalog (1FGL) operating in the 100 MeV - 100 GeV range \citep{Abdo10b} and $>$ 10 GeV \citep{Ackermann13}.
While $\gamma$-ray AGN are preferentially found among blazars \citep{Mattox01,Abdo10,Ackermann11},
the properties of PKS\,0202$+$149 are not fully consistent with those of the typical {\it Fermi}-detected blazars.
At radio frequencies the {\it Fermi}-detected $\gamma$-ray blazars indicate a core dominated flux and a higher degree of polarization than the blazars that are not detected \citep{Chen14,Jorstad10}.
However, the core component of PKS\,0202$+$149 only accounts for about 50\% of the integrated flux density measured by the Effelsberg 100 m telescope \citep{Bondi96} and has a low ($p < 1\%$) degree of polarization.
Although $\gamma$-ray emission has been predicted in hot spots and lobes of a compact symmetric object (CSO) and compact symmetric source (CSS) \citep{Stawarz08,Kino07,Kino09,Kino11}, unambiguous evidence of $\gamma$-ray emission from CSOs and CSSs remains limited to a few possible candidates, such as 0954$+$556 \citep{McConville11}, PMN J1603$-$4904 \citep{Muller14}, 2234$+$282 \citep{An16} and 3C~286 \citep{Ackermann15}.

The classification of blazars relates to a geometrical orientation of the nuclear jet and does not define the nature of the underlying host galaxy. Therefore, the radio classification of PKS\,0202$+$149 as a
CSS
source with a strong nuclear component resulting from its jet oriented close to the line-of-sight is of interest,  because it would confirm the presence of a class of AGN sources with $\gamma$-ray emission that do not exhibit most extreme blazar properties.
In this paper we revisit the radio classification of PKS\,0202$+$149 and its properties relating to the general $\gamma$-ray AGN population, by means of a comprehensive analysis of all available high-resolution radio data at multiple epochs.

The structure of the paper is arranged as follows:
\autoref{sec2} introduces the data used for the analysis;
\autoref{sec3} describes the detailed radio properties of the source, including the overall continuum spectrum, the morphology at different resolutions, and the kinematics of the VLBI jet components;
\autoref{sec4} discusses the complex radio source structure and the source classification;
a summary is given in \autoref{sec5}.

In the cosmological model with H$_0 = 71$ km s$^{-1}$ Mpc$^{-1}$, $\Omega_{\rm M} = 0.27$ and $\Omega = 0.73$ \citep{Spergel07} used throughout this paper, 1 mas angular size corresponds to a 5.38 pc projected linear size at $z = 0.405$ \citep{Wright06}.


\section{Description of the data}
\label{sec2}

\subsection{MERLIN data}

0202$+$149 was observed with the MERLIN array at 5 GHz on 12 November 1998.
Five telescopes participated in this observation: Cambridge, Darnhall, Defford, Tabley and Knockin.
The total time spent on 0202+149 is about 10 hours.
The data were recorded in fourteen frequency channels each of 1 MHz width.
The total bandwidth is 14 MHz.
The data reduction were conducted in the NRAO Astronomical Imaging Processing Software (AIPS) package \citep{Greisen90}.
The amplitude calibration of MERLIN data was done with the suite of D-programs \citep{Thomasson86} at Jodrell bank observatory.
The final images were created using the MAPPLOT program in the Caltech software package.
The image parameters are listed in \autoref{tab:obs}. An image of 0202$+$149 has been presented in \autoref{fig2}-a.

\subsection{VLBA data}

The source 0202$+$149 was observed with the Very Long Baseline Array (VLBA) of the US as a part of the
EGRET-detected AGN monitoring campaign (BH065, PI: X.-Y. Hong) on 2000 February 17 at a wavelength of 18 cm (1.67 GHz) in snapshot mode.
The VLBA data were recorded with 32 MHz dual polarization bandwidth and were correlated at the NRAO (Socorro, NM, the US) with a 2-second integration time.
A map of the VLBA data has been presented in \autoref{fig2}-c and a combined image of the VLBA and MERLIN 1.66 GHz data has been presented in \autoref{fig2}-b with contour levels given in \autoref{tab:obs}.

The source 0202$+$149 is also one of the targets of AGN jet monitoring program MOJAVE\footnote{\url{http://www.physics.purdue.edu/astro/MOJAVE/index.html}} \citep{Lister09a} observing with the VLBA at 15 GHz.
These  observations of 0202$+$149 include 27 individual data sets, starting from 1995 July 28 to the latest epoch 2015 August 8.
The primary data reduction of the VLBA data, including editing of bad data points, amplitude calibration of the visibilities and fringe fitting, have been done in AIPS.
Post-processing including phase and amplitude self-calibration, as well as imaging of the data have been performed with the DIFMAP software package \citep{Shepherd94}.
The mapping results and contour levels are summarised in \autoref{tab:obs} and four representative epoch images have been presented in \autoref{fig4}.

In order to quantitatively analyze the jet properties, the ($u,v$) visibility data at 27 epochs have been fitted using a series of Gaussian components in DIFMAP.
In general, an elliptical Gaussian was used for fitting the brightest component commonly defined as the core.
Circular Gaussians were used for jet features in most cases.
The model fitting results of the detected components are listed in the \autoref{tab:mfc} with their flux density, distance and position angle relative to the core position, as well as the brightness temperature of the core component.
The uncertainties of the fitted parameters were estimated considering the various possible models from the different methods and the formulae given by \citet{Fomalont99}. The uncertainties in flux density are about $\lesssim10\%$. The errors in position are $\sim1/5$ of the beam size.

\begin{figure}
\centering
\includegraphics[width=\hsize]{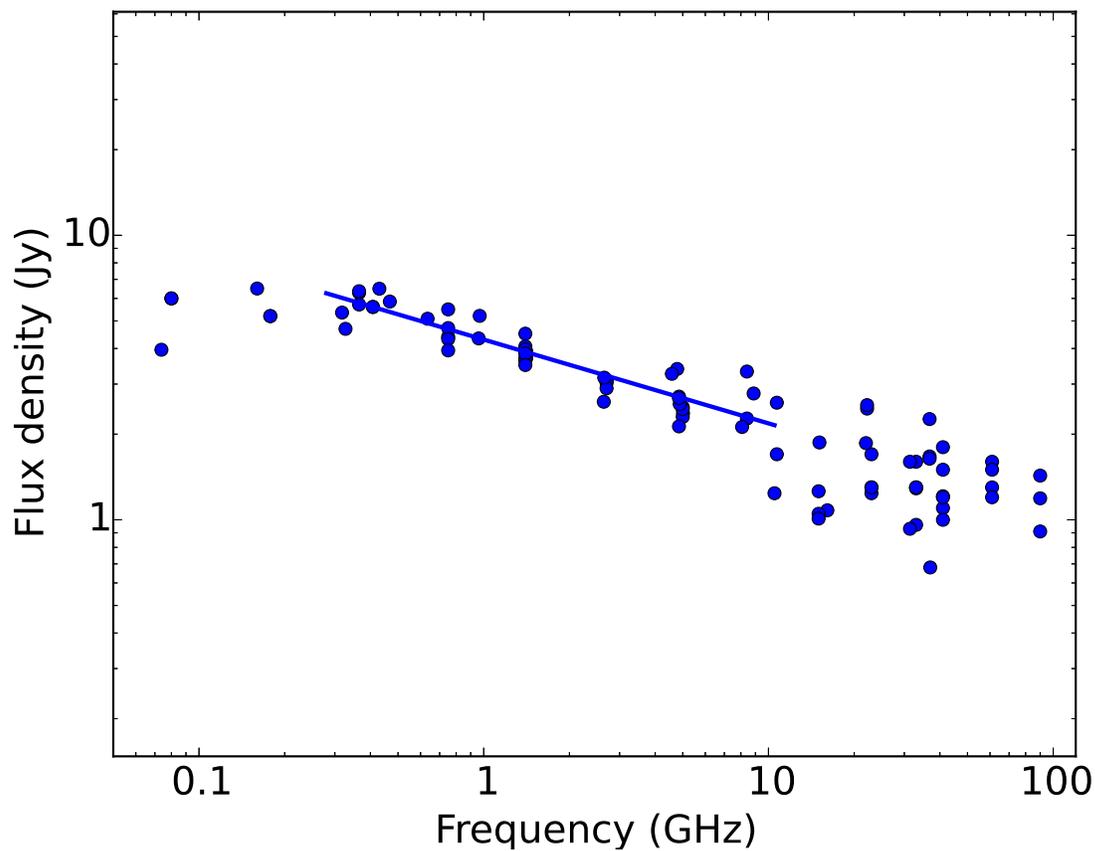}
\caption{Radio continuum spectrum from 74 MHz to 94 GHz. The flux density data are obtained from the NED. A low frequency ($\leqslant$ 300 MHz) flattening and possible turnover at $\sim 200$ MHz is indicated, aiding in our classification of this source as a CSS. A straight line power law fit between $0.3 - 10$ GHz data gives a spectral index $\alpha=0.29 \pm 0.02$. Due to the flattening at low frequencies and the possible variable flux at higher frequencies ($>$ 10 GHz), we do not use these points in the estimation of $\alpha$. For clarity purpose, the error bars of the data points are not shown.}
\label{fig1}
\end{figure}

\begin{figure*}
\centering
\includegraphics[width=1.0\hsize]{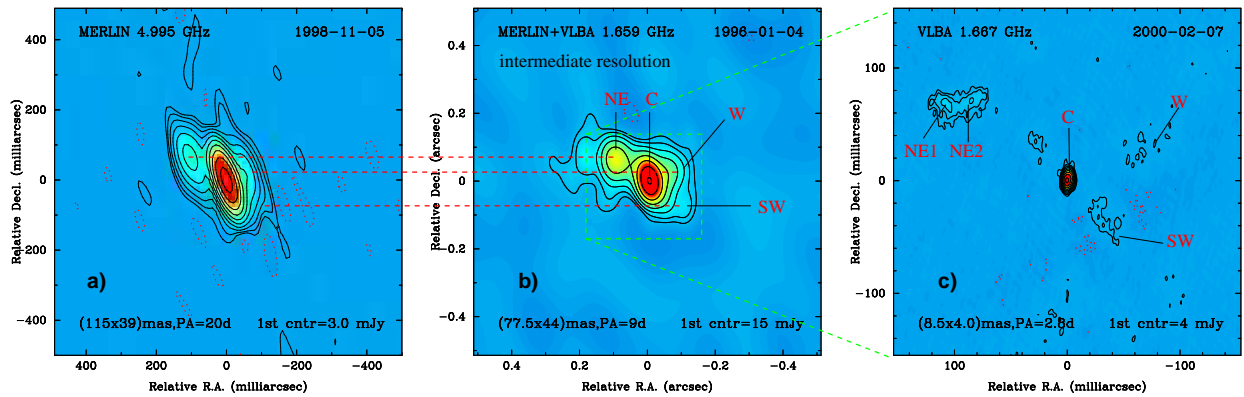}
\caption{Radio images of 0202$+$149.
{\it Left}: MERLIN image at 5.0 GHz. The image center is at RA=02$^h$04$^m$50$^s$.414, Dec=$+$15\degr14\arcmin 11\arcsec.045; {\it Middle}: an intermediate resolution
 image obtained from the combined MERLIN and VLBA data at 1.66 GHz; {\it Right}: VLBA image at 1.67 GHz.}
\label{fig2}
\end{figure*}

\begin{figure*}
\centering
\includegraphics[width=0.8\hsize]{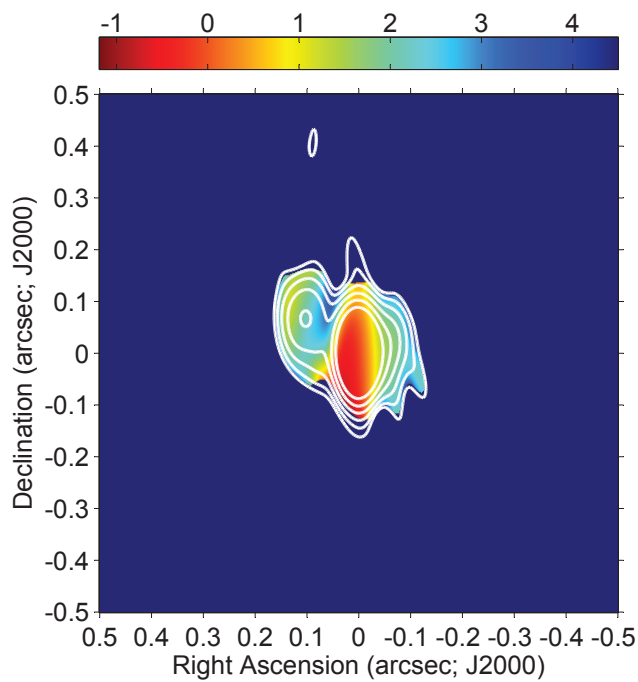}
\caption{Contour map of the spectral index $\alpha$ obtained using the MERLIN 5.0 GHz image (Figures \ref{fig2}-a) and combined MERLIN and VLBA 1.66 GHz image (Figure \ref{fig2}-b). The component C indicates an average $\alpha = 0.3$ (flat spectrum), consistent with it being the core. The NE and SW regions indicate $\alpha > 1.0$ (steep spectrum), aiding in their identification as lobes. The contours are 4 mJy beam$^{-1} \times$ (1, 2, 4, 8, 16, 32, 64). The restoring beam is 100 mas $\times$ 50 mas.}
\label{fig3}
\end{figure*}

\begin{figure*}
\centering
\includegraphics[width=0.8\hsize]{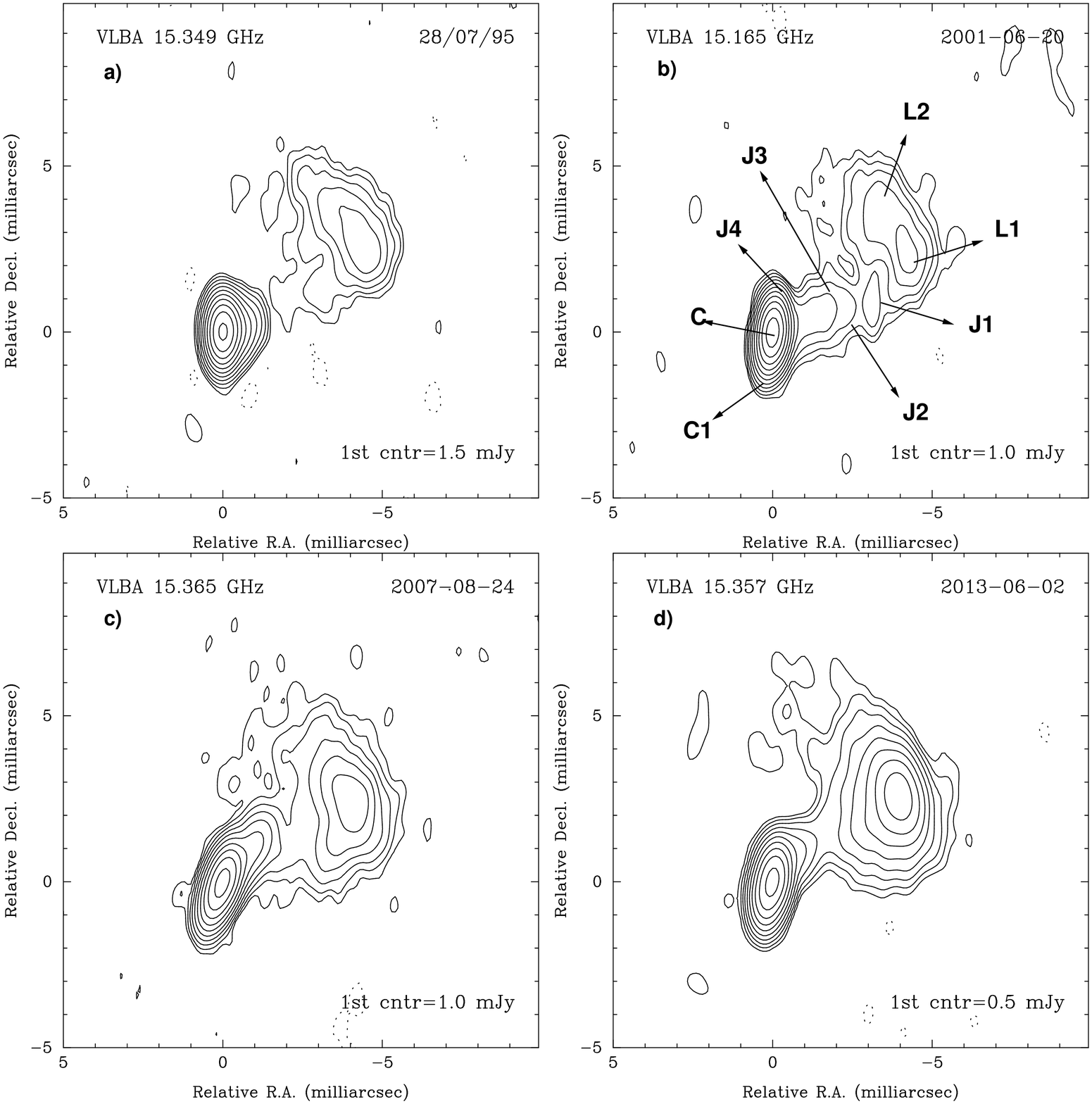}
\caption{VLBA images at 15 GHz observed at four epochs. Note the 90$\degr$ change in orientation of the one-sided jet and the related changes in the lobe structure. The structural components of Core, Jet and lobe have been identified. The contour levels are at (-1, 1, 2, 4, 8, 16, 32, 64, 128, 256, 512) $\times$ the first contour.}
\label{fig4}
\end{figure*}

\section{Results}
\label{sec3}

\subsection{Radio Continuum Spectrum}

\autoref{fig1} shows the radio continuum spectrum of the overall source, which has been determined using the flux density data of single-dish and connected interferometer measurements taken from the NED database\footnote{\url{http://ned.ipac.caltech.edu/}}.
The spectrum can be well fitted with a power law with a spectral index of $\alpha = 0.29 \pm 0.02$ ($S_\nu \propto \nu^{-\alpha}$)
between 0.3 and 10 GHz, consistent with the previous studies \citep[e.g.,][]{Herbig92}.
The spectrum of 0202+149 is slightly steeper than that of the normal blazars whose spectral index is nearly zero, for example, \citet{Fan10} obtained an averaged value of $\alpha = -0.042$ for BL Lacs and $\alpha = 0.123$ for flat-spectrum radio quasars.
This is indicative that a significant fraction of the total emission is contributed by a steep-spectrum component. The high frequency data points, which were not used in the spectrum fitting, show large scattering, probably resulting from variability  as all NED data were not obtained at the same epoch.
At low frequencies below 0.3 GHz the spectrum shows further flattening, and a hint of a turnover at about 200 MHz, a typical signature of a CSS 
object.

\subsection{The kpc- and pc-scale structures}

The morphology of the radio structure of 0202$+$149 at different frequencies and resolutions is presented in Figures \ref{fig2} and \ref{fig4}.
The image in the left-panel observed with the MERLIN at 5 GHz is characterized by a triple structure extending along northeast-southwest direction: a central compact component and two symmetric extended ones.
The total extent, {\it i.e.} the separation between NE and SW, is $\sim$250 mas, corresponding to a projected linear size of 1.3 kpc. The overall source size ($\sim$1 kpc) and the core-double-lobe morphology are reminiscent of a Medium-sized Symmetric Object (MSO, with size of 1$-$15 kpc), a population of compact double-lobed radio sources which are thought to be young radio galaxies \citep{An12}.

The integrated flux density of the emission structure revealed by the MERLIN in Figure \ref{fig2}-a is 2.49 Jy, accounting for 80 percent of the single-dish measurement obtained from the UMRAO monitoring program \citep{Aller85} at a similar frequency of 4.8 GHz on 1998 November 27.
The source displays a weak (maximum of $\sim 16\%$) and slow variability at 4.8 GHz over a time scale of about 1 year.
The largest angular scale to which the interferometer is sensitive (detectable) depends on the length of the shortest projected baseline. In this MERLIN observation, Darnhall-Tabley contributes the shortest projected baseline of 14 km, resulting in a maximal mappable size of about 0.38 arcsec.
As the MERLIN and single-dish measurements were not simultaneous (three weeks difference), the 20\% flux density discrepancy could be due to variability. 
However, 
it could also contain a contribution from the extended emission which is larger than 0.3 arcsec (the total extent in Figure \ref{fig2}-a) and resolved by the MERLIN.

The middle-panel Figure \ref{fig2}-b shows an image derived from combined MERLIN and VLBA data at 1.66 GHz.
The 1.66-GHz image has a nearly equivalent resolution with the 5-GHz MERLIN image.
Figs. \ref{fig2}-a and \ref{fig2}-b reveal consistent structure: a central component labeled as C, a northeast component NE and southwest component SW, as well as an western extension W.
We note that the earlier 1.7-GHz observations, lacking sufficient resolution and sensitivity, only detected the components NE and C \citep{Bondi96,Padrielli86,Romney84}.
Combining Figs. \ref{fig2}-a and \ref{fig2}-b, we obtained the spectral index map presented in Figure \ref{fig3}, in which C exhibits flat spectral index (an average value is $\alpha = 0.3$) and is consistent with the identification as the core. The NE and SW regions indicate a steep spectrum ($\alpha_{\rm NE} \sim 2.5$, $\alpha_{\rm W} \sim 1.0$) and are identified as lobes.

The NE and SW lobes are highly resolved and only the core remains unresolved in the 1.67-GHz VLBA image in right-panel of Figure \ref{fig2}-c with a resolution of 8.5 mas $\times$ 4.0 mas, which is ten times higher than in Fig. \ref{fig2}-a and Fig. \ref{fig2}-b.
Compared with the single-dish flux density, the VLBA accounts for 60\% of the total flux at 1.67 GHz, in agreement with the previous report, suggesting that a significant fraction is contributed by extended emission \citep{Bondi96}. The flux domination by the 0202$+$149 VLBI core, compared to the extended jet emission is rather small among {\it Fermi}-detected quasars \citep{Chen14} and supports the alternative CSS classification mentioned in Section 1.

The high resolution morphology of 0202$+$149 at 15 GHz in four representative epochs shows the presence of a nuclear component, a connecting jet structure and an inner lobe structure only on the west side of the source (Fig. \ref{fig4}). While the location of the core component and the inner lobe remain stationary, the intervening jet structure changes continuously over the course of the twenty years of monitoring. The source exhibits a global decrease in power that can be attributed to the core component. As will be shown in a later section, the launching direction of the jet changes within a 8.3$^o$ cone which suggest some instability-driven jet ejections.

\subsection{The light curve of the source}

Besides the structural variations in the core, the intervening jet structure, and the lobe, the flux of these three components varies as well. Figure \ref{fig5} top panel shows the light curve obtained from the relatively coarse sampling of the 27 MOJAVE epochs. Two nuclear outbursts may be identified in 2002 and 2008. In addition, a prominent outburst may have occurred as well in 1995, just before the monitoring started, and there is a follow-up enhancement peaking in 1999 that can be attributed to the jet plus lobe. These outbursts will be connected to distinct jet outflow events in the following section.

\subsection{Modeling the jet structure}

To quantitatively investigate the kinematics of the core-jet structure and the lobe, the emission structure at each of the 27 MOJAVE program epochs has been fitted with
a total of
six Gaussian components starting at the core C, following the jet structure and reaching into the lobe, as seen in Fig. \ref{fig4}.
An elliptical Gaussian has been used to represent the brightest core component. Circular Gaussians have been used for the jet components in order to emphasise the proper motions and changes of the emission peak with time. 
The identification of jet components has been conducted by evaluating the results at neighbouring epochs: those with consistent flux density, separation from the core, and position angle, are cross-identified as 
the
same component.

\begin{figure}
\centering
\includegraphics[width=1.1\hsize]{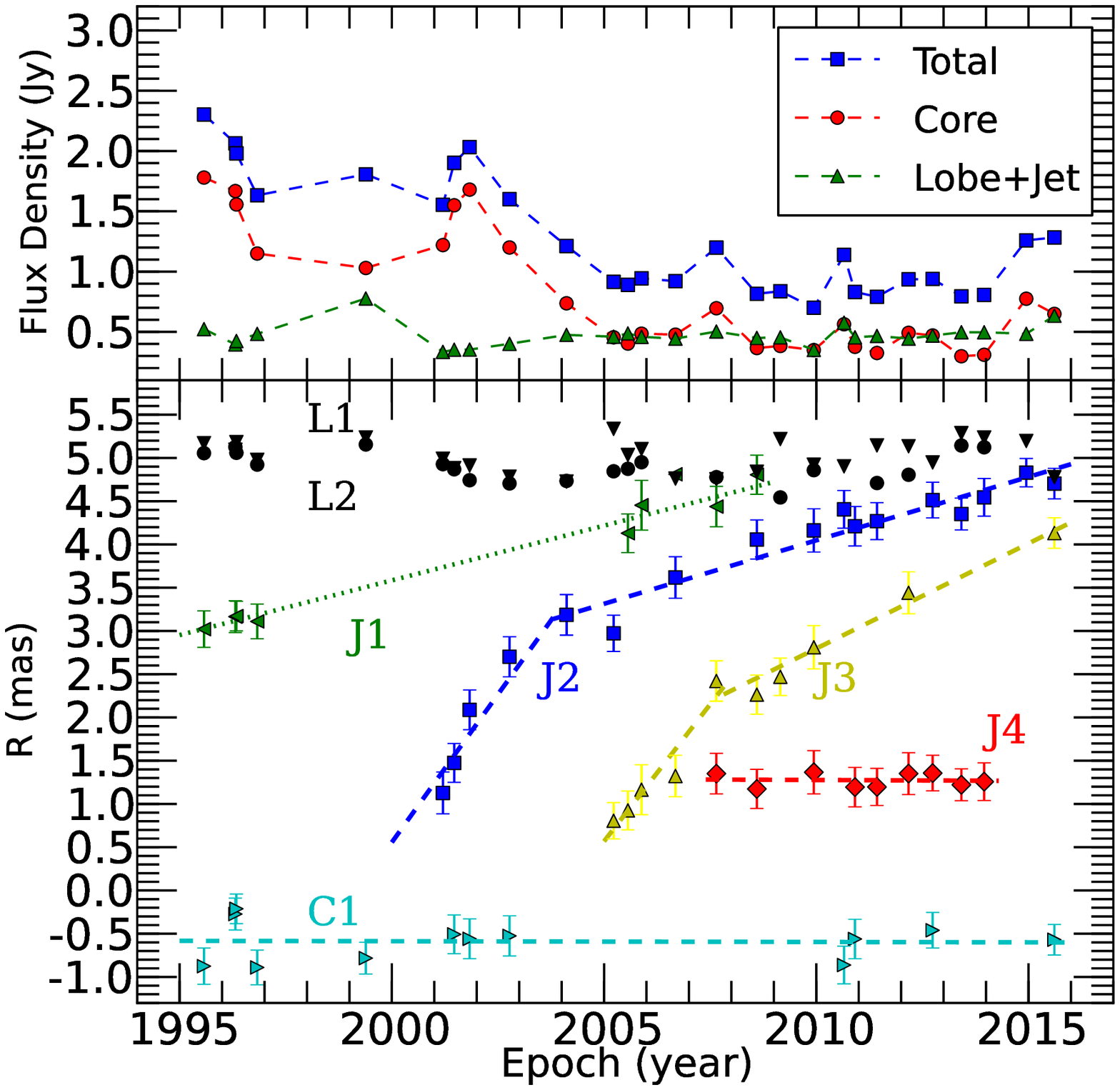}
\caption{Variations of the sources 0202+149. (a) The light curve of the source components using the 27 MOJAVE epoch images. The integrated flux density, the measured (peak) core flux density, and the differential lobe flux density are depicted. (b) Separation of the structural core-jet and lobe components presented as a function of time relative to the core component. Jet, lobe and core components have been identified as J1$-$J4, L1$-$L2 and C and C1. The error bar in the figure is 1/5 of the major axis beamsize.
}
\label{fig5}
\end{figure}

Figure \ref{fig5}-b shows the distances of the jet and lobe components relative to component C as a function of time.
The components depicted in this diagram reveal distinct structural components associated with the core C and with the variable and extended lobe structures L1 and L2. Two distinct curved jet outflows, identified as J2 (2001$-$2014) and J3 (2005$-$2015), propagate all the way between core and lobe. These components are fitted with a dashed line for analysis of their proper motions as discussed below. In addition, there is an incomplete jet component J1 during the early monitoring period (1996) that shows only halfway between C and L1 and there is a fledgling component J4 seen during the later epochs (2007$-$2015). 
It should be noted that the current mathematical modelling results are distinctly different from those resulting from that done using 10 components, presented in \citep{Lister13}.

A comparison of the two frames in Figure \ref{fig5} shows that the start of the outflows J2 and J3 clearly follows an outburst at the nucleus. In the case of J1 this connection is less clear because only a few data points of the jet outflow are recorded in the lower sensitivity maps of the early monitoring epochs. The underdeveloped jet outflow J4 also coincides with nuclear activity but a clear connection between core activity and jet event is not evident.

\begin{figure}
\centering
\includegraphics[width=1.1\hsize]{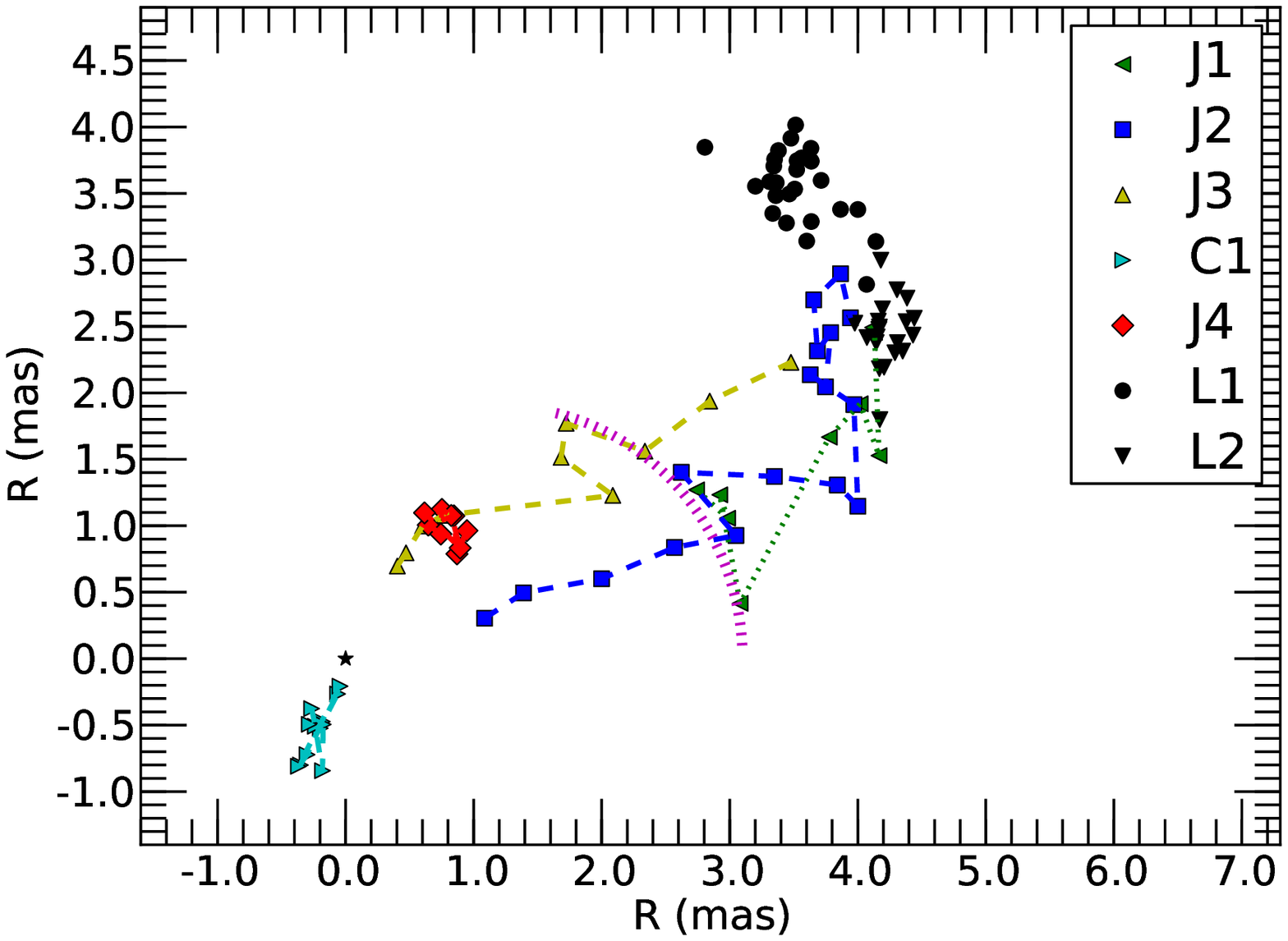}
\caption{The evolution of spatial positions of jet components as a function of time relative to the core component.
The different tracks of  jet components J1 to J4 have been indicated as well as the different entry points of J2 and J3 into the lobe.
The dashed arc indicates the region at a projected distance of 15 pc where the proper motions of the jet components are lowered and where J2 and J3 change directions.}
\label{fig6}
\end{figure}

The spatial distribution of all fitted components is presented in Figure \ref{fig6}.  
When conducting model fitting, we have noticed that the brightness peak (i.e., the core position) shows slight offset ($\lesssim$0.1 mas) from the image center. 
These position shifts result from both the addition of flux components during the launch of the jet and from the appearance of 
a
counter component at different epoch maps.
The counter component, labelled as C1 in Figure \ref{fig4}-b, lies to the south-east of the nucleus, and is consistent with evidence of counter-jet activity associated with the observed nuclear outbursts (see Table \ref{tab:mfc}).

The component data points in Figure \ref{fig6} also identify the different paths of the jet outflows J2 and J3 as well as the location of the components of the `stagnated' jet outflow J4. Jet outflow J2 is ejected in westerly direction at PA = $-80\degr$, while J3 is ejected in northwesterly direction at PA = $-20\degr$ and turns into a westerly direction PA = $-70\degr$ after a distance of 1.5 mas (8 pc). As a result of their different orientations and different paths, the entry points for the two jet outflows into the lobe structure are different, which results in a shift in the brightness distribution of the lobe with time and changes the orientation of the global structure of the jet (see Figure \ref{fig4}).

The change in direction of the jet outflows at the core suggests some kind of instability in the launching region. Since jet outflows J2 and J3 define the upper and lower boundaries of the large-scale jet region, the period of the instability would be on the order of twice the time interval between the two ejections, or about 10 years. It is not yet evident whether this would be a kink or a helical instability. Further monitoring of this source over a long time period may shed light on this question.

The 15 GHz maps presented in Figure \ref{fig4} show a (perpendicular) widening of the jet structure in a region at a radial distance of approximately 2.5 mas (15 pc), which is indicated in Figure \ref{fig6} as a dashed arc. Coincidentally this is also the distance where the proper motion decreases in J2 and J3. In addition, this is also the location of the early data points from J1 and the additional points from 2001. This distance represents a transition region in the jet flow, possibly because of the presence of a decelerating shock or a transition from supersonic to subsonic flow. This transition would result in a brightening of the structure and a decrease of the proper motion.

The dynamics and the changing characteristics of the jet outflows and the lobe structure may be understood by placing the inner lobe (L1+L2) in front of the core source and the jet outflows being foreshortened and pointing towards the observer. However, the northwest pc-scale jet and lobe structures shows a large misalignment with the kpc-scale jet (northeast-southwest orientation) indicating a change in direction by more than 90 degrees.
Since the structural link between the pc- and kpc-scale structures (beyond 5 mas) is missing in the images, one can only guess their physical connection, if they are indeed connected.  The most likely outflow pattern would connect the inner jet and the south-to-north pc-scale lobe to the northeast kpc-scale lobe.

\subsection{Proper motions of jet components}



Proper motions for three jet outflows can be determined from `radial distance-time' diagram of Figure \ref{fig5}-b.
The proper motion of this incomplete J1 outflow can only be estimated by assuming the data points from 1996 are connected to data point close to the lobe between 2005 and 2009. A dotted line based on this premise has been added to Figure \ref{fig5}-b, which suggests a proper motion of $0.13$ mas yr$^{-1}$, equivalent to superluminal motion with $\beta_{app} = 3.1\,c$.

The proper motions of the J2 and J3 outflow components have two distinct values. For J2 these proper motions are 0.68 and 0.15 mas yr$^{-1}$, which is equivalent to $\beta_{app} = 16.3\,c$ for the first part and $3.6\,c$ for the second part of the track.
Similarly for J3, the proper motions are 0.63 and 0.24 mas yr$^{-1}$ or 15.1 and $5.7\,c$ for the first and second parts of the track.
No significant proper motions can be detected in the C1 and J4 components with estimated values of 0.0008 and $-$0.002 mas yr$^{-1}$, respectively.


Previous investigations using the 8-GHz RDV (Research \& Development --- VLBA) data \citep{Piner98} resulted in non-detections of proper motions, which may be attributed to lower resolution and a relatively shorter time range of the datasets used.
Later \citet{Piner12} reported the presence of jet components with significant proper motions using an elongated time baseline. Our identification of jet components J1, J2, and J3 are different from those presented earlier.
\citet{Lister13} also estimated the jet proper motions based on the MOJAVE data. 
The proper motion speeds obtained by their group are 0.644 mas yr$^{-1}$ for J2 and 0.568 mas yr$^{-1}$ for J3.
Our calculations of J2 and J3 in their first parts of tracks show good consistency with their results. 
Since the data points used in \citet{Lister13} stopped at epoch 2010, the decreasing of jet speeds in the second part was not detected in their work.



In summary, the observational data do reveal apparent superluminal motion of components in the inner jet of 0202$+$149 combined with a stationary inner lobe structure.
The superluminal velocities of the inner part of the path of the outflows J2 and J3 are 
{\rm consistent with}
the average values in $\gamma$-ray AGN \citep[$\sim15\,c$: ][]{Jorstad01,Lister09b}.

\subsection{Core brightness temperature}

The brightness temperature of the core component is a measurement of the compactness of the radio emission structure, and is also an indicator of relativistic beaming.
The calculated $T_{\rm b}$ values of the core at 15 GHz as obtained from the component fitting (Column 8 Table \ref{tab:mfc}) are in the range of (0.1 -- 7.2) $\times 10^{11}$ K with an average of $2.0 \times 10^{11}$ K.
This value is fairly low for a radio core of a radio-loud AGN, 65\% of which have $T_{\rm b} > 10^{12}$K  \citep{Horiuchi04,Kovalev05,Moellenbrock96}.

Since the brightness temperature depends only on the physical length of the maximum projected baseline and on the core flux density \citep{Kovalev05}, the use of equivalent baseline lengths as for previous experiments indicates that the observed relatively lower value for $T_{\rm b}$ of 0202$+$149 is intrinsic.
A lower core flux density may result from a smaller intrinsic jet power or a smaller beaming factor for a jet pointing away from the line of sight.
Alternatively, the core is mixed with optically-thin plasma that is not resolved at 15 GHz, which reduces the compactness of the core and causes a lower observed brightness temperature.
This latter argument is confirmed by VLBA observations at higher frequency of 22 GHz and at higher resolution, which gave a high radio core brightness temperature of $>10^{12}$K \citep{Moellenbrock96}, and by higher-resolution 22 and 43 GHz VLBI archival data from the Radio Reference Frame Image Database\footnote{\url{http://rorf.usno.navy.mil/rrfid.shtml}}.

 The observed brightness temperature may be used to estimate the Doppler boosting factor during jet-launching stages as it is the ratio of the observed and the equipartition brightness temperature, $\delta = T_{\rm b,obs}/T_{\rm b,eq}$.  The equipartition value is a conventional and reasonable estimate of the intrinsic brightness temperature $T_{\rm b,int}$ of a compact AGN core expressed as  $T_{\rm b,eq} \simeq 5 \times 10^{10}$ K \citep{Readhead94}.
In the case of 0202+149, $\delta$ varies between 0.2 and 14.5, with an average value of 7.3. Together with the apparent superluminal speeds of 16.3 $c$ and 15.1 $c$ during the early stages of J2 and J3, an estimate of the maximum Lorentz factor of 21.9 is inferred from $\Gamma = (\beta^2_{\rm app} + \delta^2 + 1)/(2 \delta)$.  This maximal value is clearly higher than the average $\Gamma = 17.4$ deduced for the flat-spectrum radio quasar population \citep{Ghisellini93}.
Based on these results the viewing angle (see Equation B7 of Ghisellini et al. 1993) between the jet axis and the line of sight is estimated to be in the range of 3.9\degr{} and 7.0\degr{}, indicating that the inner jet aligns close to the line of sight.

\section{Discussion}
\label{sec4}

The AGN source 0202$+$149 shows properties that are a little different from many other $\gamma$-ray-emitting blazars. It shows an apparent proper motion speed that is similar with the nominal blazar values but a slightly lower brightness temperature and Doppler boosting factor \citep{Jorstad01}.
In addition, 0202+149 shows a complicated radio structure from pc to kpc scales. The kpc-scale overall structure and the flattening of the spectrum $<$ 300 MHz indicates a CSS with a central core and two low-brightness lobes. The inner pc-scale structure displays a compact core component and a one-sided curved jet that varies with time and terminates in (or bends into) an inner lobe. The core dominates the total flux density and has a brightness temperature of $\sim 10^{11}$ K, which suggests that the inner part of the pc-scale jet is relativistic and lies at a moderate viewing angle. The relativistic beaming factor would vary with the changing viewing angle of the launching jet outflows.

The pc-scale jet appears to be obstructed or deflected at two locations. A deceleration of the jet flow happens at a projected distance of 15 pc and again at the wandering entry point of the jet into the extended lobe at $\sim 25$ pc northwest of the core.
For a viewing angle $\theta = 7\degr$, the jet enters the lobe at a radial distance of about 200 pc away from the core; for an angle of only $\sim 3.9\degr$ the entry point would be at a radial distance of 370 pc.
Since a pressure-driven flow from beyond the inner lobe will seek the direction of steepest (pressure) descent, the connection between the pc-scale jet and lobe structure and the misaligned ($\sim 90\degr$) kpc-scale structure remains uncertain. The bend to the north of the inner lobe could indicate the presence of a still existing `flow channel' that leads (in front of the core source) towards the northeast kpc-scale lobe. Alternatively, the western pc-scale jet and lobe could naturally connect to the western relic component (W in Figure \ref{fig2}). Rather than observing the CSS in the plane of the sky, the kpc- and pc-scale components are superposed to give this complicated and misaligned structure.
If these outer kpc-scale lobes lie along the axis of the galaxy, they would also define the plane of the galaxy and suggest that the recent jet ejection happens close to the plane of the disk of the host galaxy, which would explain the presence of the inner lobe.


The large misalignment between the kpc and pc-scales and the absence of intermediate structures suggest that 0202$+$149 is a re-started CSS, where the kpc-scale lobes correspond to relics from past AGN activity and the inner pc-scale jet and lobe are associated with the recent AGN activity. However, the morphology of the western inner lobe suggest that this recent activity is also obstructed by some obstacle at the lobe or that the lobe represents a change in direction of the larger scale outflow channel.
Assuming that the hot spot advance speed has an average value of $0.15 \,c$ for a young radio source\citep{An12,An12b,Kawakatu06}, the distance of the outer northeast lobe of $\sim$ 0.75 kpc would suggest a kinematic age of $1.5 \times 10^{4}$ yr, consistent with the typical CSS time scale.
The re-started source may appear differently depending on whether the new jet follows the same channel paved by the previous ones or not.
Relics of past nuclear activities have also been observed in other young radio sources \citep{An12,An12b}.
Without evidence of an interaction with a nearby galaxy or a cluster environment for 0202$+$149, there would be no re-alignment of the jet in this CSS-sized object and the restarted jet may well re-fill the flow channels of the previous period of activity.

An alternative explanation of the misalignment may be a geometric projection of a helical structure of the jet itself \citep{Conway93}, as suggested by the changing position angle and the curved appearance of the pc-scale jet. After initial ejection of the forward jet, a small directional change along its helical trajectory would be significantly magnified by the projection, resulting in a large curvature from the mas to arcsec scale. Helical jets may be driven by Kelvin-Helmholz hydrodynamic instabilities of the sub-mas jet flow \citep{Camenzind86,Hardee87}, which is observed as oscillatory jet morphology in radio-loud AGNs, such as Markarian 501 \citep{Conway95}, 3C~345 \citep{Steffen95}, 1156$+$295 \citep{Hong04} and 3C~48 \citep{An10}.
Since 0202$+$149 does exhibit the characteristics differing from a blazar, the observed properties favour the CSS interpretation.

Questions still remain regarding the absence of an eastern pc-scale lobe. While the forward beaming of the western jet may explain the prominence of the intervening structure, the western lobe is not beamed. Similarly an eastern lobe should be present if the source has a two-sided jet. A time delay between the eastern and western lobe could be one explanation that requires an activity timescale less than only $1.8 \times 10^3$ yr for an estimated minimum projection angle of 5\degr.
Alternatively absorption of an eastern lobe on the other side of the nucleus may make the eastern lobe disappear, but an absorbing optical depth of 4 is required and may be unlikely at observing frequencies of 15 GHz. Therefore, the eastern lobe should be present but stacking the available data does not (yet) reveal an eastern inner lobe structure.

\section{Summary and conclusion}
\label{sec5}

The archival VLBA and MERLIN data of 0202$+$149 with multiple resolutions and at multiple frequencies reveal a complicated curved jet and lobe structure.
The radio observations of 0202$+$149 exhibit some different properties from the {\it Fermi}-detected blazars.
Its overall structure of 1.3 kpc is characterized by a triple core-lobe structure but with a large misalignment of the pc- and kpc-scale jet structures.
This is not usual in most $\gamma$-ray blazars that are recognized as a core-jet on both pc and kpc scales.
The pc-scale core, jet and inner-lobe extend to a projected distance of $\sim$25 pc to the northwest at a position angle perpendicular to the kpc-scale structure. 
Two jet components J2 and J3 exhibit superluminal motions of $\sim 16\,c$ suggesting a  relativistic inner jet pointing at a small angle to the line of sight.
The larger scale radio morphology would imply that 0202$+$149 is a CSS with the relic kpc lobes contributing about 40 \% of the total radio emission.
The nuclear source may be understood as re-started activity with a core and curved-jet structure pointing at the observer. The outflow is obstructed at a distance of about 300 pc, which results in a varying inner lobe structure and an indication of a flow channel towards the northeast relic lobe. In addition, the core brightness temperature is lower than the average value of highly-beamed $\gamma$-ray AGN. This clue confirms the presence of a mixture of optically-thick and optically-thin plasma in the core, which is also consistent with a young re-started radio source.

These radio properties are similar to those of $\gamma$-ray emitting CSO, 0954$+$556, which also exhibits a misaligned morphology from pc to kpc scales.
These two sources show that with a certain outflow velocity and jet alignment, blazar-like behaviour may also result in (intermittent) lower power sources classified as CSOs and CSSs.
 0202$+$149 provides another good case study of the $\gamma$-ray emission mechanism in lower radio power AGN, together with sources 0954$+$556, PMN J1603$-$4904 and 2234$+$282.


\section*{Acknowledgments}
This work was supported by the 973 program (2013CB837900), National Science Foundation of China (U1331205). WAB has been supported as a Visiting Professor of the Chinese Academy of Sciences (KJZD-EW-T01). PM thanks the support through the CAS PIFI program. The National Radio Astronomy Observatory is a facility of the National Science Foundation operated under cooperative agreement by Associated Universities, Inc. This research has made use of data from the MOJAVE database that is maintained by the MOJAVE team \citep{Lister09a}. The e-MERLIN is a National Facility operated by the University of Manchester at Jodrell Bank Observatory on behalf of the UK Science and Technology Facilities Council (STFC). This research has made use of the United States Naval Observatory (USNO) Radio Reference Frame Image Database (RRFID).


\begin{table*}
\centering
\scriptsize
\caption{Image parameters in Figures 1, 2 and 4.}
\begin{tabular}{lclcccc}
\hline \hline
Figure  & Epoch & Array&  Band  &  $S_{\rm peak}$ & Contours     & Beam \\
        & (yr)  &      & (GHz)  &  (Jy b$^{-1}$) & (mJy b$^{-1}$)& (mas$\times$mas, $\degr$)\\
\hline
 Fig. \ref{fig2}a&  1998.11& MERLIN       &  5.0&  2.133   & 3.0$\times$($-$1,1,2,$\ldots$,512)& 115$\times$39, 20 \\
 Fig. \ref{fig2}b&  1996.01& MERLIN$+$VLBA& 1.66&  2.001   &15.0$\times$($-$1,1,2,$\ldots$,128)& 77.5$\times$44.0, 9.0\\
 Fig. \ref{fig2}c&  2000.10& VLBA         & 1.67&  1.227   & 4.0$\times$($-$1,1,2,$\ldots$,256)& 8.5$\times$4.0, 2.8\\
 Fig. \ref{fig4}a&  1995.07& VLBA         & 15.0&  $1.78$  & 1.5$\times$($-$1,1,2,$\ldots$,512) & 1.09$\times$0.57, 0.0\\
 Fig. \ref{fig4}b&  2001.06& VLBA         & 15.0&  $1.55$  & 1.0$\times$($-$1,1,2,$\ldots$,512) & 1.09$\times$0.57, $-3.6$\\
 Fig. \ref{fig4}c&  2007.08& VLBA         & 15.0&  $0.70$  & 1.0$\times$($-$1,1,2,$\ldots$,512) & 1.09$\times$0.57, $-15.1$\\
 Fig. \ref{fig4}d&  2013.06& VLBA         & 15.0&  $0.30$  & 0.5$\times$($-$1,1,2,$\ldots$,512) & 1.09$\times$0.57, 1.2\\
\hline
\end{tabular}
\\[0.1cm]
 \label{tab:obs}
\end{table*}

\capstartfalse
\begin{deluxetable}{cccccccc}
\tabletypesize{\scriptsize}
\tablecaption{Model fitting results of 27 epochs. \label{tab:mfc}}
\tablewidth{0pt}
\tablehead{
\colhead{Epoch} &
\colhead{Comp.} &
\colhead{S$_{int}$} &
\colhead{$R$} &
\colhead{P.A.} &
\colhead{$\theta_{maj}$} &
\colhead{$\theta_{min}$} &
\colhead{T$_b$} \\
\colhead{} &
\colhead{} &
\colhead{(Jy)} &
\colhead{(mas)} &
\colhead{($\degr$)} &
\colhead{(mas)} &
\colhead{(mas)} &
\colhead{($10^{11}$K)} \\
\colhead{(1)} &
\colhead{(2)} &
\colhead{(3)} &
\colhead{(4)} &
\colhead{(5)} &
\colhead{(6)} &
\colhead{(7)} &
\colhead{(8)}
}
\startdata
1995.073  &C  &  1.771 	 &  0.0 	 &  0.0   	 & 0.28 	 & 0.09 	 & 5.3  \\
          &   &  0.144 	 &  0.43 	 & $-$85.2   & 0.32 	 & 0.32 	 &     \\
          &C1 &  0.023 	 &  0.88 	 & 156.2 	 & 1.67 	 & 1.67 	 &     \\
          &J1 &  0.023 	 &  3.02 	 & $-$65.1 	 & 1.67 	 & 1.67 	 &     \\
          &L1 &  0.106 	 &  5.17 	 & $-$45.9 	 & 1.33 	 & 1.33 	 &     \\
          &L2 &  0.104 	 &  5.06 	 & $-$61.2 	 & 0.80 	 & 0.80 	 &     \\
\hline
1996.042  &C  &  1.550 	 &  0.0    	 & 0.0     	 & 0.22 	 & 0.07 	 & 7.2  \\
          &   &  0.185 	 &  0.42 	 & $-$57.9 	 & 0.41 	 & 0.41 	 &     \\
          &   &  0.017 	 &  0.97 	 & $-$66.1 	 & 0.36 	 & 0.36 	 &     \\
          &C1 &  0.078 	 &  0.27 	 &   166.7 	 & 0.01 	 & 0.01 	 &     \\
          &J1 &  0.008 	 &  3.16 	 & $-$70.5 	 & 0.52 	 & 0.52 	 &     \\
          &L1 &  0.106 	 &  5.09 	 & $-$43.7 	 & 1.43 	 & 1.43 	 &     \\
          &L2 &  0.106 	 &  5.12 	 & $-$60.0 	 & 0.88 	 & 0.88 	 &     \\
\hline
1996.050  &C  &  1.520 	 &  0.0    	 & 0.0    	 & 0.26 	 & 0.09 	 & 4.8  \\
          &   &  0.153 	 &  0.46 	 & $-$63.7 	 & 0.36 	 & 0.36 	 &     \\
          &   &  0.028 	 &  0.86 	 & $-$72.0 	 & 0.70 	 & 0.70 	 &     \\
          &C1 &  0.047 	 &  0.21 	 &   167.6 	 & 0.25 	 & 0.25 	 &     \\
          &J1 &  0.014 	 &  3.17 	 & $-$66.9 	 & 1.11 	 & 1.11 	 &     \\
          &L1 &  0.104 	 &  5.18 	 & $-$43.2 	 & 1.54 	 & 1.54 	 &     \\
          &L2 &  0.108 	 &  5.05 	 & $-$59.8 	 & 0.96 	 & 0.96 	 &     \\
\hline
1996.103  &C  &  1.169 	 &  0.0    	 & 0.0    	 & 0.26 	 & 0.09 	 & 3.7  \\
          &   &  0.051 	 &  0.47 	 & $-$82.4 	 & 0.10 	 & 0.10 	 &     \\
          &   &  0.024 	 &  0.79 	 & $-$84.1 	 & 0.28 	 & 0.28 	 &     \\
          &C1 &  0.007 	 &  0.88 	 &   155.5 	 & 0.16 	 & 0.16 	 &     \\
          &J1 &  0.045 	 &  3.11 	 & $-$82.3 	 & 0.79 	 & 0.79 	 &     \\
          &L1 &  0.076 	 &  4.97 	 & $-$44.8 	 & 0.98 	 & 0.98 	 &     \\
          &L2 &  0.086 	 &  4.92 	 & $-$61.1 	 & 1.38 	 & 1.38 	 &     \\
\hline
1999.053  &C  &  1.075 	 &  0.0    	 & 0.0    	 & 0.27 	 & 0.07 	 & 4.0  \\
          &   &  0.404 	 &  0.52 	 & $-$48.2 	 & 0.10 	 & 0.10 	 &     \\
          &   &  0.201 	 &  0.74 	 & $-$60.0 	 & 0.19 	 & 0.19 	 &     \\
          &   &  0.054 	 &  1.07 	 & $-$59.8 	 & 0.46 	 & 0.46 	 &     \\
          &C1 &  0.006 	 &  0.22 	 &   157.4 	 & 0.19 	 & 0.19 	 &     \\
          &L1 &  0.061 	 &  5.24 	 & $-$43.4 	 & 1.21 	 & 1.21 	 &     \\
          &L2 &  0.065 	 &  5.16 	 & $-$59.9 	 & 0.98 	 & 0.98 	 &     \\
\hline
2001.032  &C  &  1.179 	 &  0.0 	 &  0.0 	   & 0.22 	 & 0.09 	 & 4.6  \\
          &   &  0.027 	 &  0.38 	 & $-$84.0 	 & 0.10 	 & 0.10 	 &     \\
          &J2 &  0.054 	 &  1.13 	 & $-$74.3 	 & 1.01 	 & 1.01 	 &     \\
          &N1 &  0.022 	 &  2.83 	 & $-$72.2 	 & 1.24 	 & 1.24 	 &     \\
          &L1 &  0.085 	 &  4.99 	 & $-$42.1 	 & 1.59 	 & 1.59 	 &     \\
          &L2 &  0.049 	 &  4.93 	 & $-$61.8 	 & 0.90 	 & 0.90 	 &     \\
\hline
2001.062  &C  &  1.520 	 &  0.0 	 &  0.0	     & 0.20 	 & 0.08 	 & 6.8  \\
          &   &  0.049 	 &  0.33 	 & $-$100.3	 & 0.13 	 & 0.13 	 &     \\
          &C1 &  0.004 	 &  0.51 	 &    159.0	 & 1.00 	 & 1.00 	 &     \\
          &J2 &  0.042 	 &  1.48 	 & $-$70.4 	 & 1.00 	 & 1.00 	 &     \\
          &N1 &  0.021 	 &  3.09 	 & $-$73.5 	 & 1.28 	 & 1.28 	 &     \\
          &L1 &  0.072 	 &  4.88 	 & $-$42.7 	 & 1.47 	 & 1.47 	 &     \\
          &L2 &  0.049 	 &  4.87 	 & $-$61.8 	 & 0.82 	 & 0.82 	 &     \\
\hline
2001.103  &C  &  1.502 	 &  0.0 	 & 0.0    	 & 0.40 	 & 0.12 	 & 2.4 \\
          &   &  0.037 	 &  0.37 	 & $-$108.4	 & 0.04 	 & 0.04 	 &     \\
          &C1 &  0.005 	 &  0.56 	 &    159.3	 & 1.64 	 & 1.64 	 &     \\
          &J2 &  0.045 	 &  2.09 	 & $-$73.7 	 & 1.64 	 & 1.64 	 &     \\
          &L1 &  0.080 	 &  4.91 	 & $-$42.9 	 & 1.67 	 & 1.67 	 &     \\
          &L2 &  0.051 	 &  4.74 	 & $-$62.9 	 & 0.86 	 & 0.86 	 &     \\
\hline
2002.101  &C  &  1.069 	 &  0.0    	 & 0.0    	 & 0.35 	 & 0.14 	 & 1.6  \\
          &   &  0.140 	 &  0.24 	 & $-$89.4 	 & 0.10 	 & 0.10 	 &     \\
          &   &  0.014 	 &  1.02 	 & $-$110.8  & 0.33 	 & 0.33 	 &     \\
          &C1 &  0.173 	 &  0.52 	 & 160.5 	 & 0.01    & 0.01    &    \\
          &J2 &  0.049 	 &  2.71 	 & $-$72.0 	 & 1.31 	 & 1.31 	 &     \\
          &L1 &  0.069 	 &  4.78 	 & $-$42.0 	 & 1.52 	 & 1.52 	 &     \\
          &L2 &  0.058 	 &  4.71 	 & $-$62.4 	 & 0.94 	 & 0.94 	 &     \\
\hline
2004.021  &C  &  1.024 	 &  0.0 	 & 0.0 	     & 0.35 	 & 0.12 	 & 1.8 \\
          &   &  0.020 	 &  0.96 	 & $-$140.2	 & 0.20 	 & 0.20 	 &          \\
          &J2 &  0.041 	 &  3.12 	 & $-$71.8 	 & 1.19 	 & 1.19 	 &          \\
          &L1 &  0.069 	 &  4.79 	 & $-$43.5 	 & 1.47 	 & 1.47 	 &          \\
          &L2 &  0.058 	 &  4.71 	 & $-$62.4 	 & 0.94 	 & 0.94 	 &     \\
\hline
2005.032  &C  &  0.441 	 &  0.0 	 & 0.0 	     & 0.35 	 & 0.13 	 & 0.7  \\
          &J2 &  0.022 	 &  3.21 	 & $-$69.8 	 & 1.00 	 & 1.00 	 &          \\
          &J3 &  0.266 	 &  1.03 	 & $-$29.6 	 & 0.36 	 & 0.36 	 &          \\
          &L1 &  0.057 	 &  5.19 	 & $-$42.2 	 & 1.39 	 & 1.39 	 &          \\
          &L2 &  0.106 	 &  4.85 	 & $-$60.0 	 & 0.94 	 & 0.94 	 &          \\
\hline
2005.072  &C  &  0.406 	 &  0.0 	 & 0.0 	     & 0.40 	 & 0.14 	 & 0.5  \\
          &J1 &  0.095 	 &  4.41 	 & $-$64.8 	 & 1.57 	 & 1.57 	 &          \\
          &J3 &  0.241 	 &  1.17 	 & $-$30.0 	 & 0.37 	 & 0.37 	 &          \\
          &L1 &  0.067 	 &  5.17 	 & $-$41.4 	 & 1.56 	 & 1.56 	 &          \\
          &L2 &  0.059 	 &  5.00 	 & $-$57.6 	 & 0.55 	 & 0.55 	 &          \\
\hline
2005.112  &C  &  0.462 	 &  0.0 	 & 0.0 	     & 0.58 	 & 0.22 	 & 0.3  \\
          &J1 &  0.087 	 &  4.45 	 & $-$65.4 	 & 1.50 	 & 1.50 	 &          \\
          &J3 &  0.262 	 &  1.16 	 & $-$31.9 	 & 0.36 	 & 0.36 	 &          \\
          &L1 &  0.056 	 &  5.10 	 & $-$41.8 	 & 1.42 	 & 1.42 	 &          \\
          &L2 &  0.072 	 &  4.95 	 & $-$58.6 	 & 0.61 	 & 0.61 	 &          \\
\hline
2006.091  &C  &  0.554 	 &  0.0 	 & 0.0 	     & 0.94 	 & 0.29 	 & 0.2 \\
          &   &  0.056 	 &  0.38 	 & $-$5.1    & 0.10 	 & 0.10 	 &     \\
          &J1 &  0.153 	 &  4.81 	 & $-$58.8 	 & 1.09 	 & 1.09 	 &     \\
          &J2 &  0.031 	 &  3.62	 & $-$67.7 	 & 2.34 	 & 2.34 	 &     \\
          &J3 &  0.082 	 &  1.32 	 & $-$35.4 	 & 0.32 	 & 0.32 	 &     \\
          &L1 &  0.051 	 &  4.76 	 & $-$36.1 	 & 3.36 	 & 3.36 	 &     \\
\hline
2007.082  &C  &  0.855 	 &  0.0 	 & 0.0 	     & 0.75 	 & 0.25 	 & 0.3 \\
          &J1 &  0.061 	 &  4.43 	 & $-$69.9 	 & 1.37 	 & 1.37 	 &     \\
          &J3 &  0.041 	 &  2.42 	 & $-$69.5 	 & 1.30 	 & 1.30 	 &     \\
          &J4 &  0.046 	 &  1.35 	 & $-$44.5 	 & 0.20 	 & 0.20 	 &     \\
          &L1 &  0.081 	 &  4.75 	 & $-$46.4 	 & 1.09 	 & 1.09 	 &     \\
          &L2 &  0.087 	 &  4.78 	 & $-$60.2 	 & 0.74 	 & 0.74 	 &     \\
\hline
2008.081  &C  &  0.511 	 &  0.0 	 & 0.0    	 & 0.96 	 & 0.34 	 & 0.1 \\
          &J1 &  0.097 	 &  4.81 	 & $-$59.3 	 & 0.82 	 & 0.82 	 &     \\
          &J2 &  0.078 	 &  4.06 	 & $-$71.2 	 & 1.66 	 & 1.66 	 &     \\
          &J3 &  0.040 	 &  2.26 	 & $-$48.0 	 & 0.89 	 & 0.89 	 &     \\
          &J4 &  0.019 	 &  1.17 	 & $-$47.8 	 & 0.17 	 & 0.17 	 &     \\
          &L1 &  0.064 	 &  4.84 	 & $-$44.0 	 & 1.02 	 & 1.02 	 &     \\
\hline
2009.023  &C  &  0.386 	 &  0.0 	 & 0.0 	     & 0.30 	 & 0.11 	 & 0.9  \\
          &   &  0.142 	 &  0.84 	 & $-$28.0 	 & 0.24 	 & 0.24 	 &     \\
          &J3 &  0.063 	 &  2.47 	 & $-$44.2 	 & 1.41 	 & 1.41 	 &     \\
          &L1 &  0.082 	 &  5.22 	 & $-$44.2 	 & 1.05 	 & 1.05 	 &     \\
          &L2 &  0.136 	 &  4.54 	 & $-$66.7 	 & 1.56 	 & 1.56 	 &     \\
\hline
2009.121  &C  &  0.459 	 &  0.0 	 & 0.0 	     & 0.90 	 & 0.34 	 & 0.1 \\
          &J2 &  0.036 	 &  4.16 	 & $-$73.7 	 & 1.00 	 & 1.00 	 &     \\
          &J3 &  0.037 	 &  2.84 	 & $-$55.9 	 & 1.00 	 & 1.00 	 &     \\
          &J4 &  0.017 	 &  1.37 	 & $-$37.9 	 & 0.28 	 & 0.28 	 &     \\
          &L1 &  0.052 	 &  4.92 	 & $-$44.6 	 & 1.00 	 & 1.00 	 &     \\
          &L2 &  0.090 	 &  4.86 	 & $-$58.9 	 & 1.10 	 & 1.10 	 &     \\
\hline
2010.083  &C  &  0.827 	 &  0.0 	 & 0.0 	     & 0.87 	 & 0.32 	 & 0.2  \\
          &   &  0.072 	 &  0.20 	 & $-$158.3	 & 4.15 	 & 4.15 	 &     \\
          &C1 &  0.006 	 &  0.87 	 & 167.5 	 & 0.98 	 & 0.98 	 &     \\
          &L1 &  0.074 	 &  4.89 	 & $-$47.9 	 & 1.62 	 & 1.62 	 &     \\
          &J2 &  0.401 	 &  4.41 	 & $-$64.3 	 & 1.90 	 & 1.90 	 &     \\
\hline
2010.113  &C  &  0.487 	 &  0.0 	 & 0.0 	     & 0.72 	 & 0.20 	 & 0.2  \\
          &C1 &  0.095 	 &  0.54 	 & 154.4 	 & 0.70 	 & 0.70 	 &     \\
          &J2 &  0.222 	 &  4.21 	 & $-$59.3 	 & 1.74 	 & 1.74 	 &     \\
          &J4 &  0.029 	 &  1.19 	 & $-$32.3 	 & 0.92 	 & 0.92 	 &     \\

\hline
2011.061  &C  &  0.376 	 &  0.0 	 & 0.0 	     & 0.57 	 & 0.07 	 & 0.7  \\
          &   &  0.100 	 &  0.72 	 & $-$32.2 	 & 0.12 	 & 0.12 	 &     \\
          &J2 &  0.163 	 &  4.27 	 & $-$61.8 	 & 1.98 	 & 1.98 	 &     \\
          &J4 &  0.029 	 &  1.19 	 & $-$38.5 	 & 0.15 	 & 0.15 	 &     \\
          &L1 &  0.050 	 &  5.14 	 & $-$43.4 	 & 0.95 	 & 0.95 	 &     \\
          &L2 &  0.061 	 &  4.71 	 & $-$57.9 	 & 0.68 	 & 0.68 	 &     \\
\hline
2012.030  &C  &  0.502 	 &  0.0 	 & 0.0 	     & 0.29 	 & 0.08 	 & 1.6  \\
          &   &  0.069 	 &  0.66 	 & $-$29.6 	 & 0.16 	 & 0.16 	 &     \\
          &J3 &  0.045 	 &  3.44 	 & $-$55.9 	 & 1.00 	 & 1.00 	 &     \\
          &J4 &  0.035 	 &  1.35 	 & $-$33.8 	 & 0.39 	 & 0.39 	 &     \\
          &L1 &  0.105 	 &  5.13 	 & $-$48.9 	 & 0.96 	 & 0.96 	 &     \\
          &L2 &  0.149 	 &  4.80 	 & $-$59.9 	 & 0.91 	 & 0.91 	 &     \\
\hline
2012.093  &C  &  0.520 	 &  0.0 	 & 0.0 	     & 0.57 	 & 0.17 	 & 0.4  \\
          &   &  0.041 	 &  0.49 	 & $-$32.2 	 & 0.10 	 & 0.10 	 &     \\
          &C1 &  0.018 	 &  0.46 	 & 145.4 	 & 0.11 	 & 0.11 	 &     \\
          &J2 &  0.219 	 &  4.51 	 & $-$57.1 	 & 1.83 	 & 1.83 	 &     \\
          &J4 &  0.019 	 &  1.36 	 & $-$37.8 	 & 0.39 	 & 0.39 	 &     \\
          &L1 &  0.096 	 &  4.94 	 & $-$55.3 	 & 0.56 	 & 0.56 	 &     \\
\hline
2013.061  &C  &  0.432 	 &  0.0 	 & 0.0 	     & 0.81 	 & 0.24 	 & 0.2  \\
          &J2 &  0.153 	 &  4.35 	 & $-$62.4 	 & 1.53 	 & 1.53 	 &     \\
          &J4 &  0.011 	 &  1.22 	 & $-$47.3 	 & 0.23 	 & 0.23 	 &     \\
          &L1 &  0.055 	 &  5.14 	 & $-$44.7 	 & 0.97 	 & 0.97 	 &     \\
          &L2 &  0.126 	 &  5.28 	 & $-$56.4 	 & 0.60 	 & 0.60 	 &     \\
\hline
2013.122  &C  &  0.307 	 &  0.0  	 & 0.0 	     & 0.19 	 & 0.05 	 & 2.4  \\
          &   &  0.068 	 &  0.78 	 & $-$23.6 	 & 0.14 	 & 0.14 	 &     \\
          &J2 &  0.130 	 &  4.54 	 & $-$53.5 	 & 2.10 	 & 2.10 	 &     \\
          &J4 &  0.020 	 &  1.26 	 & $-$29.2 	 & 0.18 	 & 0.18 	 &     \\
          &L1 &  0.072 	 &  5.23 	 & $-$49.8 	 & 0.56 	 & 0.56 	 &     \\
          &L2 &  0.067 	 &  5.12 	 & $-$57.2 	 & 0.56 	 & 0.56 	 &     \\
\hline
2014.121  &C  &  0.878 	 &  0.0 	 & 0.0	     & 0.35 	 & 0.35 	 & 2.3  \\
          &   &  0.047 	 &  0.97 	 & $-$28.0 	 & 0.27 	 & 0.27 	 &     \\
          &J2 &  0.239 	 &  4.83 	 & $-$52.0 	 & 1.73 	 & 1.73 	 &     \\
          &L1 &  0.139 	 &  5.19 	 & $-$53.3 	 & 0.62 	 & 0.62 	 &     \\
\hline
2015.122  &C  &  0.550 	 &  0.0 	 &0.0 	     & 0.29 	 & 0.29 	 & 0.5  \\
          &C1 &  0.080 	 &  0.57 	 &150.1	     & 0.81 	 & 0.81 	 &     \\
          &J2 &  0.165   &  4.70 	 &$-$57.0	 & 2.09 	 & 2.09 	 &     \\
          &J3 &  0.134 	 &  4.13 	 &$-$57.3	 & 0.59 	 & 0.59 	 &     \\
          &L1 &  0.348 	 &  4.78 	 &$-$48.9	 & 0.06 	 & 0.06 	 &     \\
\enddata
\tablenotetext{(1) }{the observation epoch;}
\tablenotetext{(2) }{the component label;}
\tablenotetext{(3) }{the flux density of each component;}
\tablenotetext{(4), (5) }{the distance and position angle of each component with respect to core;}
\tablenotetext{(6), (7) }{the major and minor axis of the Gaussian component;}
\tablenotetext{(8) }{the brightness temperature of the core.}
\end{deluxetable}
\capstarttrue

\end{document}